\documentclass[pre,aps,amsmath,amssymb,reprint,floatfix,superscriptaddress]{revtex4-1}

\pdfoutput=1
\usepackage{xcolor}
\usepackage{siunitx}
\usepackage{graphicx}
\usepackage[colorlinks=true,citecolor=blue,urlcolor=blue]{hyperref}
\usepackage{etoolbox} 
\usepackage{bm} 
\usepackage[normalem]{ulem}
\newcommand{\new}[1]{\textcolor{black}{#1}}

\newcommand{\thin}{\mskip0.75mu}

\def\eqq#1{Eq.~(\ref{#1})}

\def\eq#1{(\ref{#1})}

\def\f#1{Fig.~\ref{#1}}

\def\c#1{~\cite{#1}}

\def\cc#1{~Ref.~\cite{#1}}

\def\av#1{\langle #1 \rangle}

\def\beq{\begin{equation}}
\def\eeq{\end{equation}}
\def\bea{\begin{eqnarray}}
\def\eea{\end{eqnarray}}

\def\x{{\bm x}}
\def\tt{{\bm \theta}}

\def\nv{N_{\rm v}}
\def\nh{N_{\rm h}}

\def\kB{k_{\rm B}}
\def\tf{t_{\rm f}}
\def\kt{\kB T}

\begin{document}

\title{Generative thermodynamic computing}

\author{Stephen Whitelam}
\email{swhitelam@lbl.gov}

\affiliation{Molecular Foundry, Lawrence Berkeley National Laboratory, 1 Cyclotron Road, Berkeley, CA 94720, USA}

\begin{abstract}

We introduce a generative modeling framework for thermodynamic computing, in which structured data is synthesized from noise by the natural time evolution of a physical system governed by Langevin dynamics. While conventional diffusion models use neural networks to perform denoising, here the information needed to generate structure from noise is encoded by the dynamics of a thermodynamic system. Training proceeds by maximizing the probability with which the computer generates the reverse of a noising trajectory, which ensures that the computer generates data with minimal heat emission. We demonstrate this framework within a digital simulation of a thermodynamic computer. If realized in analog hardware, such a system would function as a generative model that produces structured samples without the need for artificially-injected noise or active control of denoising.

\end{abstract}

\maketitle

{\em Introduction ---} In this paper we describe a generative modeling framework for thermodynamic computing. Thermodynamic computing is closely related to the field of probabilistic computing\c{kaiser2021probabilistic,aadit2022massively,misra2023probabilistic}. It is based on the ideas that we can do energy-efficient computation by using small physical devices whose microscopic states change with time in response to thermal fluctuations, and that the fluctuations of a suitably-designed device can encode the outcome of a desired calculation\c{conte2019thermodynamic,hylton2020thermodynamic,wimsatt2021harnessing,aifer2024thermodynamic,melanson2025thermodynamic}. Here we show that thermodynamic computers can perform generative modeling in a manner analogous to diffusion models, with key differences.

In a diffusion model, structured inputs are degraded by the gradual addition of noise. A neural network is trained to enact the reverse process, allowing the generation of structure from noise\c{sohl2015deep,yang2023diffusion,biroli2024dynamical,yu2025nonequilbrium}. This process is implemented on a digital computer, where noise is introduced in the form of artificially generated pseudorandom numbers. Here we use analytic calculations and digital simulations to suggest an alternative approach, in which the noise-driven dynamics of a thermodynamic computer -- the noise arising naturally from the system's interaction with its environment -- generates structure from noise. If realized in analog hardware, such a system would generate structured outputs simply by evolving with time under its natural dynamics. It would not require added pseudorandom noise, or the guidance of a digital neural network. 

In more detail, we consider a model thermodynamic computer, a set of fluctuating nonlinear degrees of freedom coupled by bilinear interactions. The computer's degrees of freedom evolve according to overdamped Langevin dynamics. This design is inspired by existing hardware that can perform linear algebra\c{aifer2024thermodynamic}, and by our recent work showing that a nonlinear version of such hardware can function as the thermodynamic version of a neural network\c{whitelam2024thermodynamic}. We provide input to the computer to make it display images of digits from the MNIST data set\c{mnist_leaderboard}, and allow these images to degrade by running the dynamics of the computer with its inter-unit couplings set to zero. Such degradation is called {\em noising} in the diffusion model literature. As we do so, we compute from the Langevin equation the probability that a computer with hypothetical nonzero couplings would have generated the {\em reverse} of this noising trajectory, and we adjust the values of these hypothetical couplings by gradient descent in order to maximize that probability. After running several such noising trajectories, we construct a denoising computer using the trained couplings, and verify that its natural dynamics, starting from noisy initial conditions, leads to the generation of structured MNIST-like digits. Independent dynamical trajectories of the same computer produce a variety of outcomes, some of which are not contained in the training set.

 In this approach the denoising dynamics is encoded by the couplings of the trained thermodynamic computer, which plays the role of a denoising neural network in a diffusion model. If realized in analog hardware -- for example, using networks of mechanical\c{dago2021information}, electrical\c{melanson2025thermodynamic}, or superconducting\c{ray2023gigahertz} oscillators -- the information required for denoising would be encoded in the energy landscape of the computer, rather than in a digital neural network. As a result, denoising would not be simulated, but physically enacted. 
 
In this mode of operation the thermodynamic computer resembles a nonequilibrium, continuous-spin analog of a Boltzmann machine, a statistical mechanical model that represents probability distributions over binary variables\c{hinton2017boltzmann,salakhutdinov2009deep}. \new{The key difference is that a Boltzmann machine encodes information in its equilibrium Boltzmann distribution, whereas our device runs on a physical clock: the computation is carried out by the dynamics of the system at a designated time, with no requirement to attain equilibrium. We therefore refer to the device considered here, which uses Langevin trajectories to perform a calculation, as a {\em Langevin computer}.}

We also show that the training process has a direct physical interpretation: it adjusts the computer's couplings in order to minimize the thermodynamic irreversibility of the generative process. By finding the computer most likely to have generated the reverse of a noising trajectory, we minimize the expected heat emission and entropy production of the denoising computer. Our results therefore link the design of generative thermodynamic models to fundamental physical principles, and broaden our understanding of the capabilities of thermodynamic computers.

{\em Training a generative Langevin computer ---} Consider a model of a thermodynamic computer. The computer is composed of $N$ classical, real-valued fluctuating degrees of freedom $\x=\{x_i\}$, which could represent voltage states in electrical circuits\c{melanson2025thermodynamic}, oscillator positions in a mechanical system\c{ dago2021information}, or phases in Josephson Junction devices\c{ray2023gigahertz,pratt2025controlled}. The computer's units $x_i$ evolve in time according to the overdamped Langevin dynamics
\beq
\label{lang1}
\dot{x}_i= -\mu \thin \partial_i V_\tt(\x)  + \sqrt{2 \mu \thin \kt} \, \eta_i(t).
\eeq 
Here $\mu$ is the mobility parameter, which sets the basic time constant of the computer. For the thermodynamic computers of Refs.\c{aifer2024thermodynamic,melanson2025thermodynamic}, $\mu^{-1}$ is of order a microsecond. For damped oscillators made from mechanical elements\c{dago2021information} or Josephson junctions\c{ray2023gigahertz,pratt2025controlled}, $\mu^{-1}$ is of order a millisecond or a nanosecond, respectively. The first term on the right-hand side of \eqq{lang1} is the force arising from the computer's potential energy $V_\tt(\x)$, given a set of parameters (couplings and biases) $\tt=(\{J_{ij}\},\{b_i\})$; note that $\partial_i \equiv \partial/\partial x_i$. The second term on the right-hand side of \eqq{lang1} models temporally uncorrelated thermal fluctuations: $\kt$ is the thermal energy scale, and the Gaussian white noise terms satisfy $\av{\eta_i(t)}=0$ and $\av{\eta_i(t) \eta_j(t')} = \delta_{ij} \delta(t-t')$. 

The potential energy $V_\tt(\x)$ of the computer is
\beq
\label{pot}
V_\tt(\x) = \sum_{i=1}^N  \left(J_2 x_i^2+J_4 x_i^4\right)+\sum_{i=1}^N b_i x_i +\sum_{(ij)} J_{ij} x_i x_j.
\eeq
The first sum in \eqq{pot}, which runs over the $N$ units, sets the intrinsic couplings of the computer. For $J_4=0$ we have a linear model\c{aifer2024thermodynamic}, whose unit activations are linear functions of their inputs, while for $J_4 >0$, \new{the case we consider}, we have a nonlinear model that can act as the thermodynamic analog of a neural network\c{whitelam2024thermodynamic} \new{(positive $J_4$ also ensures the thermodynamic stability of the computer as the $J_{ij}$ are adjusted). We consider the case $J_2 >0$, which creates units with one stable state, analogous to the s-units of thermodynamic computing\c{melanson2025thermodynamic}. The alternative choice, $J_2<0$, creates bistable units, analogous to the p-spins in the field of probabilistic computing\c{kaiser2021probabilistic}.}

The remaining terms in \eq{pot} contain the trainable parameters of the computer. The parameters $b_i$ are input signals or biases applied to each unit. The parameters $J_{ij}$ are pairwise couplings between units, inspired by the bilinear interactions of the thermodynamic computers of Refs.\c{aifer2024thermodynamic,melanson2025thermodynamic}, with the sum running over all inter-unit connections. \eqq{pot} describes a thermodynamic computer of arbitrary connectivity, and the following discussion applies to the same. 

Imagine that we observe a dynamical trajectory of the computer at a series of discrete times, $\omega = \{\x(t_k)\}_{k=0}^{K}$, where $t_k = k \Delta t$. The probability that any step of this trajectory was generated by a thermodynamic computer with parameters $\tt$ can be calculated from the Onsager-Machlup action associated with the Langevin equation\c{risken1996fokker,cugliandolo2017rules}. A time-discretized version of this action can be derived by first considering a standard Euler integration scheme for \eqq{lang1},
\beq
\label{lang2}
x_i(t+\Delta t)=x_i(t) -\mu \thin \partial_i V_\tt({\bm x})\Delta t  + \sqrt{2 \mu \thin \kt \Delta t} \, \eta_i.
\eeq 
Here $\Delta t$ is the integration timestep, and $\eta_i$ is a Gaussian random variable with zero mean and unit variance. Writing $\Delta x_i \equiv x_i(t+\Delta t)-x_i(t)$, we can rearrange \eq{lang2} to read
\beq
\label{eta}
\eta_i=\frac{\Delta x_i+\mu \thin \partial_i V_\tt(\x) \Delta t}{\sqrt{2 \mu \thin \kt \Delta t}}.
\eeq
Next, note that the probability of generating the step $\x \to \x + \Delta \x$ is that of drawing $N$ noise values $\eta_i$,
\beq
\label{step}
P^{\rm step}_\tt(\Delta \x)=(2 \pi)^{-N/2} \prod_{i=1}^N \exp(-\eta_i^2/2),
\eeq
with the $\eta_i$ given by \eqq{eta}. Hence the negative log-probability that a computer with parameters $\tt$ generated the step $\x \to \x +\Delta \x$ is 
\beq
-\ln P^{\rm step}_\tt(\Delta \x) =  \sum_{i=1}^N \frac{\left(\Delta x_i+\mu \thin \partial_i V_\tt(\x) \Delta t \right)^2}{4 \mu \thin \kt \Delta t},
\eeq
up to an unimportant constant term. The negative log-probability that a computer with parameters $\tt$ would generate the {\em reverse} step, $\x +\Delta \x \equiv \x' \to \x$, is
\beq
\label{rev}
-\ln \tilde{P}^{\rm step}_\tt(\Delta \x) =  \sum_{i=1}^N \frac{\left( -\Delta x_i+\mu \thin \partial_i V_\tt(\x') \Delta t \right)^2}{4 \mu \thin \kt \Delta t}.
\eeq
To increase the probability with which the computer would have generated the entire reverse trajectory $\tilde{\omega} = \{\x(t_{K-k})\}_{k=0}^{K}$,  we can sum \eqq{rev} over all steps of the trajectory, differentiate that expression with respect to each parameter of the computer, and update the parameters as
\bea
\label{grad1}
J_{ij} &\to& J_{ij} +\alpha \sum_{k=1}^{K} \frac{\partial} {\partial J_{ij}} \ln \tilde{P}^{\rm step}_\tt(\Delta \x(t_k)), \\
\label{grad2}
b_{i} &\to& b_{i} +\alpha \sum_{k=1}^{K} \frac{\partial}{\partial b_{i}} \ln \tilde{P}^{\rm step}_\tt(\Delta \x(t_k)) ,
\eea
where $\alpha$ is a learning rate. Recall that $\Delta \x(t_k)$ is the displacement generated at timestep $k$ in the observed (forward) trajectory. The gradient terms in \eq{grad1} and \eq{grad2} can be calculated analytically, and are
\bea
\label{grad3}
-\frac{\partial} {\partial J_{ij}} \ln \tilde{P}^{\rm step}_\tt(\Delta \x)&=&   \frac{ -\Delta x_i+\mu \thin \partial_i V_\tt(\x') \Delta t}{2 \kt } x_j \\
&+& \frac{ -\Delta x_j+\mu \thin \partial_j V_\tt(\x') \Delta t}{2 \kt } x_i \nonumber,
\eea
and
\beq
\label{grad4}
-\frac{\partial} {\partial b_i}\ln \tilde{P}^{\rm step}_\tt(\Delta \x) =   \frac{ -\Delta x_i+\mu \thin \partial_i V_\tt(\x') \Delta t}{2 \kt },
\eeq
where
\beq
\label{grad5}
\partial_i V_\tt(\x)=2 J_2 x_i+4 J_4 x_i^3+b_i+\sum_{j \in \mathcal{N}(i)} J_{ij} x_j.
\eeq
Here $\mathcal{N}(i)$ denotes the set of units connected to unit $i$.

When the forward trajectories depict noising processes, training over many such trajectories identifies couplings that allow a thermodynamic computer to transform noise into structured data.

\begin{figure} 
   \centering
   \includegraphics[width=\linewidth]{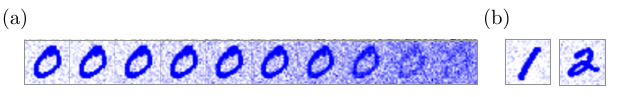} 
   \caption{(a) Example noising trajectory. (b) The remaining digits used in the training set.}
   \label{fig1}
\end{figure}

{\em Numerical illustration of generative thermodynamic computing  ---} To illustrate this result we carried out a digital simulation of a thermodynamic computer. We set $J_2=J_4 = 10 \kt$, and consider a computer with $\nv=28^2$ visible units and $\nh=512$ hidden units. The visible units will be used as a display, and the hidden units used to do computation. The computer has $\nv \nh$ trainable couplings $J_{ij}$ between visible and hidden units, and $\nh (\nh-1)/2$ trainable couplings between hidden units. Hidden units have trainable biases $b_i$. Visible units have no trainable biases, but during training we impose visible-unit biases $b_i \propto P_i$, where $P_i$ denotes the $i^{\rm th}$ pixel of an MNIST digit (each digit's pixels were adjusted to have zero mean and unit variance). We display the visible units in a $28 \times 28$ grid, matching the presentation of an MNIST digit.

To construct noising trajectories we project an MNIST image onto the visible units, via their non-trainable biases. We project part of the same image onto the hidden-unit biases, in order to provide them with some signal, and set all couplings $J_{ij}$ to zero. We then let the computer come to equilibrium, by simulating \eqq{lang1} for a sufficiently long time. We then run a dynamical trajectory of time $\tf=2.5$, slowly diminishing the intensity of the imposed digit. The result is an image that becomes increasingly noisy, as shown in \f{fig1}(a). Positive values of the unit activations $x_i$ are shown blue, while negative values are shown white. 

As we run each noising trajectory, we update Equations \eq{grad1} and \eq{grad2}. This process, repeated over many trajectories, identifies the parameters $\tt$ of the computer that would, with maximum likelihood, generate the reverse of the trajectory, and so convert noise into signal. In this small-scale example we trained the computer using only three digits shown in \f{fig1}(a) and (b).

In \f{fig2}(a) we show three independent trajectories of the thermodynamic computer trained in this way. Trajectories begin from a noisy initial state prepared by bringing the coupling-free thermodynamic computer to equilibrium. In each case, the trained thermodynamic computer gradually transforms noise into structure, illustrating its ability to perform noise-to-structure generation. These results indicate that the computer has internalized representations of the digits and can reproduce them via physical evolution.  In the case of the digit ``1'', the generated image appears inverted. This behavior reflects a breaking of the visible layer's approximate symmetry under the transformation $x_i \to -x_i$ (the digits presented to the computer have pixel values that are roughly symmetric about zero).

\f{fig2}(b) we show the output at time $t=\tf$ of 25 independent trajectories generated by the trained thermodynamic computer. The outputs exhibit diversity in style and form, behavior that is typical of diffusion models trained to approximate, rather than exactly replicate, the data distribution. Not all outputs are necessarily desirable or clearly interpretable -- some display ``mode mixing'', in the language of diffusion models -- but this simple example is intended as a proof of principle, a demonstration that structure can be generated from noise by physical dynamics alone.
\begin{figure} 
   \centering
   \includegraphics[width=\linewidth]{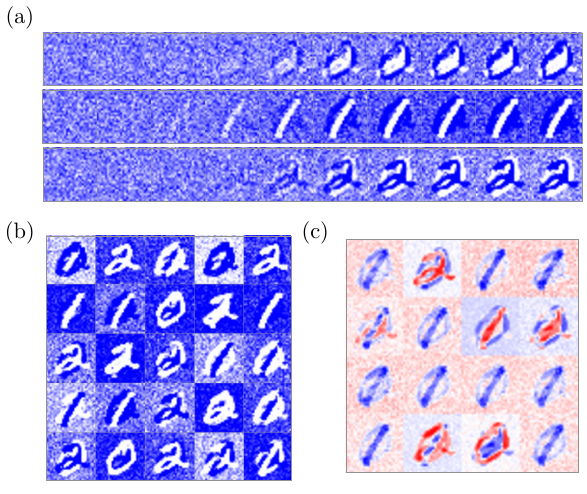} 
   \caption{(a) Three independent dynamical trajectories of the trained denoising thermodynamic computer. (b) The outcome at time $t=\tf$ of 25 independent trajectories of the trained computer. (c) Coupling patterns between 16 representative hidden units and the visible layer.}
   \label{fig2}
\end{figure}

Some insight into the behavior of the computer is provided by \f{fig2}(c), which shows 16 of the computer's hidden units' learned couplings to the visible layer. Each panel represents the pattern of couplings $J_{ij}$ between one hidden unit and all visible units: blue indicates positive coupling, red negative (either can be excitatory or inhibitory, via the interaction $J_{ij} x_i x_j$, depending on the values of the unit activations $x_i$). These patterns act as receptive fields that determine how each hidden unit responds to features in the visible layer. Several of the units have learned localized, digit-like structures, suggesting that the computer decomposes input patterns into interpretable visual components. Some hidden units exhibit complementary structure -- activating for certain strokes while suppressing others -- enabling the system to encode multiple features. Together, these receptive fields determine the energy landscape that guides the computer's dynamics, allowing the system to transform noise into structured outputs.

{\em Physical interpretation of the learning process ---} The generative model we have considered is a thermodynamic system, and we can interpret the learning algorithm used to train it in physical terms. To do so, consider one step of a noising trajectory $\omega = \{\x(t_k)\}_{k=0}^{K}$, generated by any means. Consider the ratio of two probabilities (a standard device in nonequilibrium statistical mechanics\c{seifert2012stochastic}): the probability that the step was generated by a computer with a set of reference couplings (call them $\tt = {\bm 0}$), and the probability that our denoising computer (with parameters $\tt$) generated the reverse of that step. This ratio is 
\bea
\ln \frac{P^{\rm step}_{\bm 0}(\Delta \x)}{\tilde{P}^{\rm step}_\tt(\Delta \x)} &=&  \sum_{i=1}^N \frac{\left( -\Delta x_i+\mu \thin \partial_i V_\tt(\x') \Delta t \right)^2}{4 \mu \thin \kt \Delta t} \nonumber \\
&-&  \sum_{i=1}^N \frac{\left( \Delta x_i+\mu \thin \partial_i V_{\bm 0}(\x) \Delta t \right)^2}{4 \mu \thin \kt \Delta t} \nonumber\\
&\approx&  -\sum_{i=1}^N \frac{\Delta x_i \partial_i V_{\bm 0}(\x)+\Delta x_i \partial_i V_\tt(\x) }{2 \kt} \nonumber\\
&\approx&  - \frac{\Delta Q_{\bm 0}+\Delta Q_\tt }{2 \kt}, \nonumber\\
 \eea
 to leading order in $\Delta t$. Here $\Delta Q_{\bm 0}$ and $\Delta Q_\tt$ denote the incremental heat dissipated during the forward step by the reference and trained computers, respectively. Over the entire trajectory, in the limit $\Delta t \to 0$, we have
\beq
\label{fr}
\ln \frac{P_{\bm 0}[\omega]} {P_\tt[\tilde{\omega}]}=-\frac{1}{2} (\beta Q_{\bm 0}(\omega)+ \beta Q_\tt(\omega)).
\eeq
Here $\beta \equiv 1/(\kt)$, and $P_{\bm 0}[\omega]$ and $P_\tt[\tilde{\omega}]$ denote the probabilities of observing the forward trajectory with the reference computer, and the reverse trajectory with the denoising computer, respectively. The terms $Q_{\bm 0}(\omega)$ and $Q_\tt(\omega)$ represent the total heat dissipated along the forward trajectory by the two computers.

Training proceeds by minimizing the negative log-probability \(-\ln P_\tt[\tilde{\omega}]\). According to the fluctuation relation \eq{fr}, and given that the reference process is fixed, this is equivalent to minimizing $-Q_\tt(\omega)$, the negative total heat dissipated by the denoising computer when generating the noising trajectory $\omega$. Given that heat changes sign upon time reversal, the learning process therefore minimizes the heat $Q_\tt(\tilde{\omega})=-Q_\tt(\omega)$ emitted by the trained computer along the trajectory $\tilde{\omega}$, i.e. as it generates structure from noise. In this sense, the trained dynamics is thermodynamically optimal: it is the dynamics that reconstructs the imposed data with the least heat emitted or entropy produced.

\new{{\em Thermodynamic advantage --- } We can estimate the thermodynamic advantage, the ratio of the energy costs of digital to thermodynamic computation, by considering the energy scales of denoising using a digital neural network and a hardware version of our simulated thermodynamic computer. The basic energy scale of a digital neural network is set by a multiply-accumulate (MAC) operation, which differs by hardware implementation but is typically about 1 pJ\c{horowitz20141}, or $2.4\times10^8\, \kt$ at room temperature. For a modest multilayer perceptron denoiser ($784\!\to\!128\!\to\!128\!\to\!784$), a single denoising step requires $\sim 2.2\times10^5$ MACs. If we make the very conservative assumption that the denoiser is used only 10 times within a denoising trajectory, the order-of-magnitude energy budget of denoising using a neural network is not less than $5 \times 10^{14} \, \kt$.}

\new{The energy cost of the thermodynamic computer is much smaller. We can calculate the heat emitted by the computer from the difference of the potential energy \eq{pot} between the start and end of a trajectory, $Q = V_{\tt}(\x(0))-V_{\tt}(\x(\tf))$ (no work is done on the computer after the hidden-unit biases are established). Over $1000$ independent denoising trajectories of the trained computer we calculate a mean heat emission of $\langle Q \rangle = 2.9\times10^{3}\,k_{\rm B}T$, with standard deviation $3.5\times10^{2}\,k_{\rm B}T$.  Comparing this value to the digital estimate gives a ratio of more than $10^{11}$. That is, if implemented in hardware, the thermodynamic computer would be more than ten orders of magnitude more energy efficient than a digital neural network. The example shown here is rudimentary by the standards of state-of-the-art diffusion models, but shows the potential, and potential energy savings, of thermodynamic computation.}

{\em Conclusions ---} We have proposed a generative modeling framework in which structure is produced from noise by the physical dynamics of a thermodynamic system. This approach follows the logic of a diffusion model, but instead of using a digital neural network and artificially injected noise, the information required for generation is encoded in the system's energy landscape and emerges from a physical dynamics. Used in this way, the thermodynamic computer is a Langevin computer, a nonequilibrium, continuous-spin analog of a Boltzmann machine\c{hinton2017boltzmann,salakhutdinov2009deep}.

The learning process, maximizing the likelihood that a trained system could have produced the reverse of a noising trajectory, admits a natural interpretation in terms of entropy production: the model learns to reverse the forward dynamics in the most thermodynamically reversible way. Thermodynamic learning therefore links generative modeling to fundamental physical principles. 

\cc{coles2023thermodynamic} proposed the idea of making a generative model by controlling analog physical dynamics with a digital neural network. Here we have shown that analog hardware on its own can be generative. However, having a neural network control the couplings of the thermodynamic computer would indeed make it more expressive: for instance, the neural network could adjust the couplings of the computer as a function of time, or set the computer's couplings so as to produce conditioned outputs.

We have used digital simulation to demonstrate that nonlinear, nonequilibrium analog hardware can learn to generate structured outputs from noise. This work provides another example of a thermodynamic computer trained to operate under nonequilibrium conditions\c{whitelam2024thermodynamic}. Realized physically -- trained digitally, with the learned couplings implemented in hardware -- such systems could perform autonomous generative computation without external control or artificial randomness, opening new avenues of exploration for physically grounded, energy-efficient machine learning.

{\em Acknowledgments---} I thank Isaac Tamblyn and Adrianne Zhong for discussions. This work was done at the Molecular Foundry, supported by the Office of Science, Office of Basic Energy Sciences, of the U.S. Department of Energy under Contract No. DE-AC02-05CH11231, and partly supported by US DOE Office of Science Scientific User Facilities AI/ML project ``A digital twin for spatiotemporally resolved experiments''.


%

\vspace*{-1em}
\begin{center}
\color{gray!50}
  \tiny\textit{Langevin dynamics}
  \includegraphics[width=0.9\linewidth]{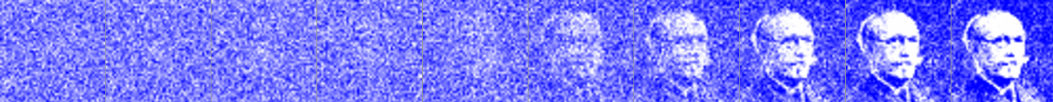}
\end{center}
\vspace*{0.5em}

\end{document}